\documentclass[twocolumn,showpacs,preprintnumbers,prl,fleqn]{revtex4}

\usepackage{graphicx}
\def\ul{    }
\begin{document}
\title{
Kondo effect in the presence of itinerant-electron ferromagnetism
  \\  studied with the numerical renormalization group method }
\author{J. Martinek,$^{1,4}$ M. Sindel,$^2$ L. Borda,$^{2,3}$ J.
Barna\'s,$^{4,5}$ J. K\"onig,$^{1,6}$ G. Sch\"on,$^1$ and J. von
Delft$^2$}
 \affiliation{$^1$Institut f\"ur Theoretische
Festk\"orperphysik, Universit\"at Karlsruhe, 76128 Karlsruhe,
Germany\\
 $^2$Sektion
Physik and Center for Nanoscience, LMU M\"unchen, Theresienstrasse
37, 80333 M\"unchen, Germany \\
 $^3$ Hungarian Academy of Sciences,
  Institute of Physics, TU Budapest, H-1521, Hungary \\
 $^4$Institute of Molecular Physics, Polish Academy of
Sciences, 60-179 Pozna\'n, Poland \\
 $^5$Department of Physics, Adam Mickiewicz University, 61-614 Pozna\'n,
 Poland \\
 $^6$Institut f\"ur Theoretische Physik III, Ruhr-Universit\"at Bochum, 44780
Bochum, Germany }

\date{Submitted: April 16, 2003}

 \begin{abstract}
The Kondo effect in quantum dots (QDs) - artificial magnetic
impurities - attached to ferromagnetic leads is studied with the
numerical renormalization group method. It is shown that the QD
level is spin-split due to presence of ferromagnetic electrodes,
leading to a suppression of the Kondo effect. We find that the
Kondo effect can be restored by compensating this splitting with
 a magnetic field. Although the resulting Kondo resonance then has
an unusual spin asymmetry with a reduced Kondo temperature, the
ground state is still a locally-screened state, describable by
Fermi liquid theory and a generalized Friedel sum rule, and
transport {\ul at zero temperature} is spin-independent.
\end{abstract}



\maketitle

{\it Introduction}. -- The prediction \cite{glazman} and
experimental observation of the Kondo effect in artificial
magnetic impurities -- semiconducting quantum dots (QDs)
\cite{goldhaber,kondo-odd} -- renewed interest in the Kondo effect
and opened new opportunities of research. The successful
observation of the Kondo effect in molecular QDs like carbon
nanotubes \cite{cobden,bachtold} and single molecules
\cite{molecular} attached to metallic electrodes opened the
possibility to study the influence of many-body correlations in
the leads (superconductivity \cite{superconducting} or
ferromagnetism) on the Kondo effect. Recently the question arose
whether the Kondo effect in a QD attached to ferromagnetic leads
can occur or not. Several authors have predicted
\cite{sergueev,zhang,bulka,lopez} that the Kondo effect should
occur. However, it was shown recently \cite{martinek1} that
 the QD level will be spin-split due to the presence of ferromagnetic
electrodes leading to a suppression of the Kondo effect, and that
the Kondo resonance can be restored only by applying an external
magnetic field. The analyses of
Refs.~\cite{sergueev,zhang,bulka,lopez,martinek1} were all based
\mbox{on approximate methods.}

In this Letter we resolve the controversy by adapting the
numerical renormalization-group (NRG) technique
\cite{costi1,hewson-book}, one of the most accurate methods
available to study strongly-correlated systems in the Kondo
regime, to the case of a QD coupled to ferromagnetic leads with
parallel magnetization directions.
We find that in general the Kondo resonance is split, similar to
the usual magnetic-field-induced splitting
\cite{hofstetter2,costi}.
 However, we find that by appropriately tuning an external magnetic
  field, this splitting can be fully compensatedand and the Kondo effect can be
  restored \cite{RKKY_vs_K} (confirming
  Ref.~\cite{martinek1}). We point out that precisely at this field
  the occupancy of the local level is the same for spin up and down,
  $n_\uparrow= n_\downarrow$, a fact that follows from the Friedel
sum rule \cite{langreth}.
 Moreover, we show that the Kondo effect
then has unusual properties such as a strong
 spin polarization of the Kondo resonance
and, just as for ferromagnetic materials,
for the density of states (DOS).
Nevertheless, despite of the spin assymetries in the
DOS of the QD and the leads,
the symmetry in the occupancy $n_\uparrow=
n_\downarrow$
implies that the system's ground state
  can be tuned to have a fully compensated local spin, in which case
  the QD conductance is found to be the \emph{same} for each spin
  channel, $ G_{\uparrow} = G_{\downarrow}$.

{\it The Model}. --
 For ferromagnetic leads electron-electron interactions in the
leads give rise to magnetic
 order and spin-dependent DOS
 $\rho_{r\uparrow}(\omega) \neq \rho_{r\downarrow}(\omega)$, $r=L,R$.
Magnetic order of typical band ferromagnets like Fe, Co, and Ni is
mainly related to electron correlation effects in the relatively
narrow $3d$ sub-bands, which only weakly hybridize with $4s$ and
$4p$ bands \cite{nolting}. We can assume that due to a strong
spatial confinement of $d$ electron orbitals, the contribution of
electrons from $d$ sub-bands to transport across the tunnel
barrier can be neglected \cite{tsymbal}.
In such a situation the system can be modeled by
noninteracting \cite{interaction}
$s$ electrons, which are spin polarized due to the exchange
interaction with uncompensated magnetic moments of the completely
localized $d$ electrons. In mean-field approximation one can
model this exchange interaction as an effective molecular field,
which removes spin degeneracy in the system of noninteracting
conducting electrons, leading to a spin-dependent DOS.
The Anderson model (AM) for a QD with a single
energy level $\epsilon_{\rm d}$, which is coupled to ferromagnetic
leads, is given by
\begin{eqnarray}
\tilde{H} &=& \sum_{rk \sigma} \epsilon_{rk
\sigma}c_{rk\sigma}^{\dagger}
  c_{rk \sigma}
  + \epsilon_{\rm d} \sum_{\sigma} \hat n_\sigma + U
  \hat n_\uparrow \hat n_\downarrow
\nonumber\\
  && + \sum_{r k \sigma} (V_{r,k} d_{\sigma}^{\dagger}   c_{r,k \sigma} +
  V^*_{r,k} c_{r,k \sigma}^{\dagger}d_{\sigma}) -
   g \mu_B B S_{z}
  \; .
  \label{eq:AMf}
\end{eqnarray}
Here $c_{rk\sigma}$ and $d_\sigma$ ($  \hat n_\sigma =
d_{\sigma}^{\dagger} d_{\sigma} $) are the Fermi operators for
electrons with momentum $k$ and spin $\sigma$ in the leads
($r=L/R$), and in the QD, $V_{rk}$ is the tunneling amplitude,
$S_{z} = (\hat n_\uparrow- \hat n_\downarrow)/2$, and the last
term is the Zeeman energy of the dot. In general, all information
about spin asymmetry in the leads can be modeled by the
spin-dependent hybridization function
$
\Gamma_{r \sigma} ( \omega ) = \pi \sum_{ k } \delta ( \omega -
\epsilon_{k \sigma} ) V_{r, k}^2
= \pi \rho_{r \sigma}(\omega) V^2_r$,
where  $V_{r, k} = {\rm const} \equiv V_r$ and
$\rho_{r \sigma}(\omega)$ is the spin-dependent DOS.

In order to understand the Kondo physics of a QD attached to two
identical ferromagnetic electrodes with parallel configurations, it
suffices to study, instead of the above general model
[Eq.~(\ref{eq:AMf})], a simpler one, which captures the same essential
physics, namely the fact that $\Gamma_{r \uparrow} ( \omega ) \neq \Gamma_{r
  \downarrow}(\omega)$ will generate an effective local magnetic
field, which lifts the degeneracy of the local level (even for $B=0$).
A simple (but not unique) way of modeling this effect is to take the DOS
in the leads to be constant and spin-independent, $\rho_{r \sigma}
(\omega) \equiv \rho$, the bandwidths to be equal $D_{\uparrow} =
D_{\downarrow}$, and lump all spin-dependence into the  spin-dependent
hybridization function, $\Gamma_{r \sigma} (\omega)$, which we take to
be $\omega$-independent, $\Gamma_{r\sigma } (\omega) \equiv \Gamma_{r
  \sigma}$.  By means of a unitary transformation \cite{glazman} the
AM [Eq.~(\ref{eq:AMf})] can be mapped onto a model in which the
correlated QD level couples only to one electron reservoir described
by Fermi operators $\alpha_{k \sigma}$ with strength $ \Gamma_{\sigma}
= \sum_r \Gamma_{r \sigma} $,
 \begin{eqnarray}\label{eq:AMf_cont}
H &=& \sum_{k \sigma} \epsilon_{k }\alpha_{k\sigma}^{\dagger}
  \alpha_{k \sigma}
  + \epsilon_{\rm d} \sum_{\sigma} \hat n_\sigma + U
 \hat  n_\uparrow \hat  n_\downarrow \\
  && + \sum_{k \sigma} \sqrt{\frac{\Gamma_{\sigma}}{\pi \rho}}
(d_{\sigma}^{\dagger}   \alpha_{k \sigma} +
  \alpha_{k \sigma}^{\dagger}d_{\sigma}) -
  g \mu_B B S_{z}
  \; . \nonumber
   \end{eqnarray}
Finally, we parametrize the spin-dependence of $\Gamma_{\sigma}$
in terms of a spin-polarization parameter $ P \equiv (
\Gamma_{\uparrow} - \Gamma_{\downarrow} ) /  \Gamma $, by writing
$\Gamma_{\uparrow ( \downarrow ) } \equiv {1 \over 2} \Gamma (1
\pm P)$, where $\Gamma \equiv \Gamma_\uparrow +
\Gamma_\downarrow$.

In the model of Eq.~(\ref{eq:AMf_cont}), we allowed for $
\Gamma_{\uparrow} \neq \Gamma_{\downarrow} $ but not for $
D_{\uparrow} \neq D_{\downarrow} $, as would be appropriate for
real ferromagnets, whose spin-up and down bands always have a
Stoner splitting $ \Delta D \equiv D_{\uparrow} - D_{\downarrow}
$,  with typical values $ \Delta D / D_{\uparrow} \lesssim 20 \% $
(for Ni, Co, Fe). However, no essential physics is thereby lost,
since the consequence of  taking $ D_{\uparrow} \neq
D_{\downarrow} $ is the same as that of taking $ \Gamma_{\uparrow}
\neq \Gamma_{\downarrow} $, namely to generate an effective local
magnetic field \cite{Stoner}.

The occurrence of the Kondo effect requires spin fluctuations in the dot
as well as zero-energy spin-flip excitations in the leads.
Indeed, a Stoner ferromagnet without full spin polarization $ -1 < P < 1 $
provides zero-energy Stoner excitations \cite{yosida}, even in the presence of
an external magnetic field.

{\it Method}. -- The model [Eq.~(\ref{eq:AMf_cont})] can be
treated by Wilson's NRG method. This method, with recent
improvements related to high-energy features and finite magnetic
field \cite{hofstetter2,costi}, is a well-established method to
study the Kondo impurity (QD) physics. It allows one to calculate
the level occupation $ n_\sigma  \equiv \langle \hat n_\sigma\rangle $ (a static property), the QD spin
spectral function, ${\rm
Im}\;\chi_s^z(\omega)=\mathcal{F}\left\{i\Theta(t)\langle
[S_z(t),S_z(0)] \rangle\right\}$, where $\mathcal{F}$ denotes the
Fourier transform, and the spin-resolved single-particle spectral
density $ A_{ \sigma}(\omega,T,B,P)  = -\frac{1}{\pi} {\rm Im}
{\cal G}_{d,\sigma}^R (\omega)$ for arbitrary temperature $T$,
magnetic field $B$ and polarization $P$ [where $G_{\sigma}^R
(\omega)$ denotes a retarded Green function]. From this we can
find the spin-resolved conductance $G_{\sigma} = {e^2 \over \hbar
} \frac{2\lambda}{(\lambda+1)^2}\Gamma_{\sigma}
 \int_{- \infty } ^{ \infty}
d \omega
A_{\sigma}(\omega) (-{ \partial f( \omega ) \over
\partial \omega } ) $
with $f( \omega )$
denoting the Fermi function and $\lambda=\Gamma_{L\sigma}/\Gamma_{R\sigma}$
describing the asymmetry in the couplings. We choose $\lambda = 1$ below.

\begin{figure}[t]
  \centerline{\includegraphics[width=8.5cm]{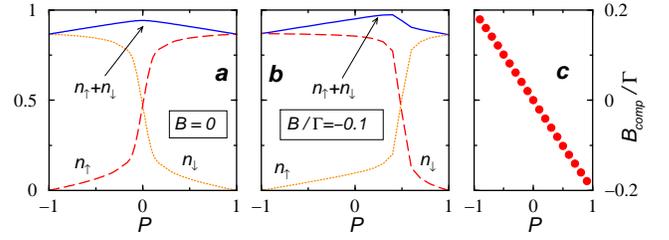}}
   \caption{
    Spin-dependent occupation of the dot level at (a) $B=0$ and
      (b) $B = -0.1 \Gamma$, as a function of spin polarization $P$.
(a) For $B=0$, the condition
 $ n_\uparrow =
 n_\downarrow $   only holds at $ P = 0 $. (b) For finite $P$ it
can be satisfied if a finite, fine-tuned magnetic field, $B_{\rm
comp} (P)$, is applied, whose dependence on $P$ is shown in (c).
As expected, it
  is approximately linear \cite{martinek1,linear}. Here $ U = 0.12 D
$, $ \epsilon_{\rm d} = -U/3 $, $ \Gamma = U/6 $, and $ T = 0$.  }
\label{fig2}
\end{figure}
\begin{figure}[t]
  \centerline{\includegraphics[width=8cm]{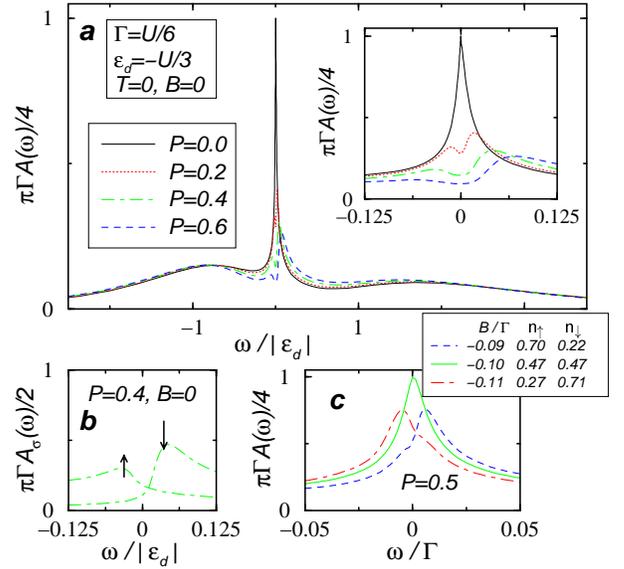}} \caption{ a) QD
    spectral function $A(\omega) = \sum_{\sigma} A_{\sigma}(\omega) $
    for several values of spin polarization $P$; inset: expanded scale
    of the spectral function around $\epsilon_{\rm F}$.  b) The
    spin-resolved spectral function for fixed $P$. For $P \to -
      P$, we have $A_{\sigma} \to A_{- \sigma}$.  c) Compensation of
    the spin splitting by fine-tuning an external magnetic field.
    Parameters $ U $, $ \epsilon_{\rm d} $, $ \Gamma $, and $T$ as in
    Fig.~\ref{fig2}.  }
\label{fig1}
\end{figure}

\begin{figure}[b]
\centerline{\includegraphics[width=8cm]{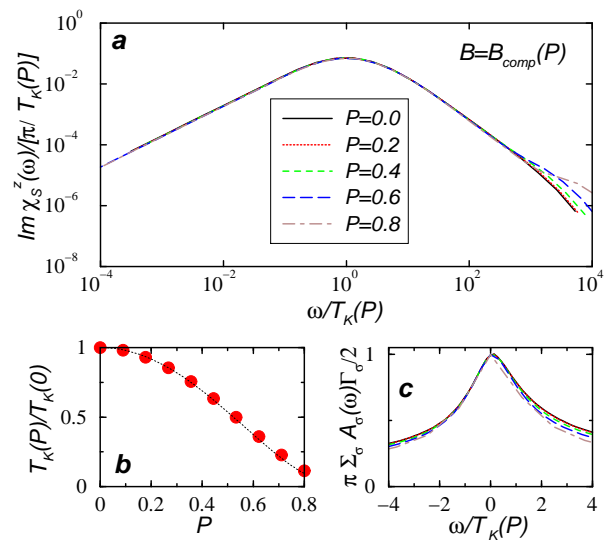}} \caption{
 a) The spin spectral function $ {\rm Im} \{ \chi_s^z(\omega) \}$
 as a function of energy.
 b) Dependence of $T_{\rm K}$ on  spin polarization $P$. The solid line
shows the prediction from Ref.~\cite{martinek1} for the functional
form $T_K(P)/T_K(0)=\exp[C \; {\rm arctanh}(P)/P ]$, where the
best fit is obtained for $C=-5.98$.
Equation~(6) from Ref.~\cite{martinek1}
with $(\rho_\uparrow+\rho_\downarrow) J_0 =
(\Gamma/\pi) U/[|\epsilon_{\rm d}|(U+\epsilon_{\rm d})]$
would lead to $C = -4.19$.
c) QD spectral function for several values of $P$.
 Parameters  $ U $, $ \epsilon_{\rm d} $, $ \Gamma $, and $ T $ are as
 in Fig.~\ref{fig2}, $B=B_{\rm comp}(P)$.} \label{fig3}
\end{figure}

{\it Generation and restoration of spin splitting}. --
 In this
Letter, we focus exclusively on the properties of the system
  at $ T = 0 $ in the local moment - Kondo regime, where the total
  occupancy of the local level, $n = \sum_\sigma n_\sigma \approx 1$.
 The occurence of charge fluctuations of course broadens and
shifts the position of the QD levels (for both spin up and down),
and hence changes their occupation. For $P \neq 0$, the charge
fluctuations and hence level shifts and level occupations become
\emph{spin-dependent}, causing the QD level to split
\cite{martinek1} and the dot magnetization $ n_\uparrow -
n_\downarrow $ to be finite (Fig.~\ref{fig2}). As a result, the
Kondo resonance, too, is spin-split \cite{hofstetter3,exception}
and weakened (Fig.~\ref{fig1}), similarly to the effect of an
applied magnetic field \cite{hofstetter2,costi}. This means that
Kondo correlations are reduced or even suppressed in the presence
of ferromagnetic leads.
 However, for any fixed $P$, it is possible
to compensate the splitting of the Kondo resonance
[Fig.~\ref{fig1}(c) and \ref{fig3}(c)] by fine-tuning the magnetic
field to
  an appropriate value, $B_{\rm comp}(P)$,
defined as the field which maximizes the height of the Kondo
resonance. This field is found to depend linearly \cite{linear} on $P$
  [Fig.~\ref{fig2}(c)] (as predicted in \cite{martinek1} for $U
  \rightarrow \infty $).
  Remarkably, we also
  find (throughout the local moment regime) that at $B_{\rm comp}$ the
  local occupancies satisfy $n_\uparrow =
  n_\downarrow$ [Fig.~\ref{fig2}(b)]. The fact that this occurs
simultaneously with the disappearance of the Kondo
  resonance splitting suggests that the local spin is fully screened
 at $B_{\rm comp}$.

\begin{figure}[t]
  \centerline{\includegraphics[width=6.8cm]{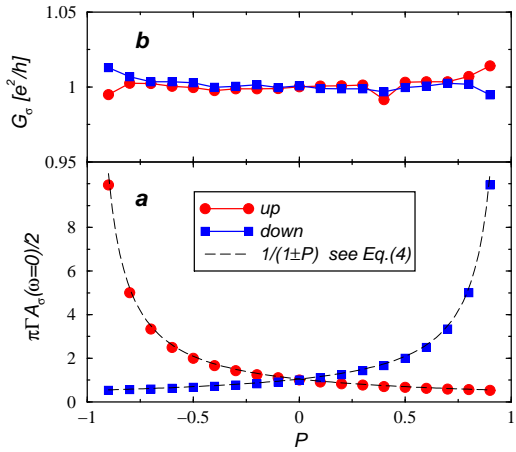}} \caption{
 (a)     Spin resolved QD spectral function amplitude
    $A_{\sigma}(\omega=0)$ at the Fermi level, and (b) the QD
    conductance $G_\sigma$,
    as functions of the spin polarization $ P $, for $B=B_{\rm
      comp}(P)$ and symmetric couplings ($\Gamma_{L \sigma} =\Gamma_{R \sigma}$), with $ U $, $
    \epsilon_{\rm d} $, $ \Gamma $, and $T$ as in Fig.~\ref{fig2},
    implying $ n_\sigma \approx 0.5 $.  The dashed line in (a) is
    $1/(1 \pm P)$ (Eq.~(\ref{eq:A}) with $n_\sigma = 0.5$).  As
    expected, we find $G_{\sigma}=e^2/h$, with a numerical error less
    than 1{\%}.
 } \label{fig4}
\end{figure}

{\it Spectral functions}. -- We henceforth fix the magnetic field
at $B = B_{\rm comp}(P)$. To learn more about the properties of
the corresponding ground state, we computed the spin spectral
function $ {\rm Im} \{ \chi_s^z(\omega) \}$ for several values of
$P$ at $T = 0$ (Fig.~\ref{fig3}). Its behavior is characteristic
for the formation of a local Kondo singlet: as a function of
decreasing frequency, the spin spectral function shows a maximum
at a frequency $\omega_{\rm max}$ which we associate with the
Kondo temperature [i.e. $k_BT_{\rm K} \equiv \hbar\omega_{\rm
max}$ at $B=B_{\rm comp}(P)$], and then decreases linearly with
$\omega$, indicating the formation of the Fermi liquid state
\cite{hewson-book}. By determining $T_{\rm K}(P)$ (from
$\omega_{\rm max}$) for different $P$-values, we find that $T_{\rm
K}$ decreases with increasing $P$ [Fig.~\ref{fig3}(b)]. For metals
like Ni, Co, and Fe, where $P = 0.24$, $0.35$, and $0.40$
respectively, the decrease of $T_{\rm K} $ is rather weak, so the
Kondo effect should still be experimentally accessible.
Remarkably, both $ {\rm Im}\{ \chi_s^z(\omega) \}$ and the
spectral function $A_{\sigma}(\omega)$ collapse rather well onto a
universal curve if plotted in appropriate units
[Figs.~\ref{fig3}(a) and \ref{fig3}(c)]. This indicates that an
applied magnetic field $B_{\rm comp}$ restores the universal
behavior characteristic for the isotropic Kondo effect, in spite
of the presence of spin-dependent coupling to the leads.
Fig.~\ref{fig4}(a) shows that the amplitude of the Kondo resonance
is strongly spin dependent for ferromagnetic leads, which is
unusual and unique. The nature of this asymmetry is related to the
asymmetry of the DOS in the leads, and its value is exactly
proportional to $ \sim 1/\Gamma_{\sigma}$. As a result the total
conductance $G_{\sigma}$ is not spin dependent
[Fig.~\ref{fig4}(b)]. This indicates the robustness of the Kondo
effect in this system: if the external magnetic field has been
tuned appropriately, it is able to compensate the presence of a
spin asymmetry in the leads by creating a proper spin asymmetry in
the dot spectral density, thereby conserving a fully compensated
local spin and achieving perfect transmission.

{\it Friedel sum rule}. --
 Further insights can be gained from
the Friedel sum rule, an exact $T=0$ relation \cite{langreth} that
holds for arbitrary values of $P$ and $B$ \cite{friedel}.
The interacting Green's function can be expressed as
\cite{hewson-book}
 $  {\cal G}^R_{d, \sigma}( \omega
) = [ \omega - \epsilon_{\rm d \sigma} + i \Gamma_{\sigma} -
\Sigma_{\sigma}(\omega) ]^{-1} $, with spin-dependent
$\epsilon_{\rm d \sigma}$ and $\Gamma_{\sigma}$; the former due to
Zeeman splitting ($ \epsilon_{\rm d \sigma} = \epsilon_{\rm d } -
1/2 \; \sigma g \mu_B B $) and latter due to the ferromagnetic
leads. Here $ \Sigma_{\sigma}(\omega ) $ denotes the
spin-dependent self energy. Now, the Friedel sum rule
\cite{langreth} implies that at $T=0$, the  occupancy $n_{\sigma}$
and spectral functions can be written as
\begin{eqnarray}
 &&  n_{\sigma} = \phi_\sigma/\pi  = {1 \over 2} -  {1 \over \pi} {\rm
tan}^{-1} \left( { \epsilon_{\rm d \sigma} -\epsilon_{\rm F} +
\Sigma_{\sigma}^{\rm R}( \epsilon_{\rm F} ) \over \Gamma_{\sigma}
} \right),  \label{eq:n} \\
 && A_{\sigma}( \epsilon_{\rm F}
)={ {\rm sin}^2(\pi  n_{\sigma} ) \over \pi \Gamma_{\sigma} } \; ,
 \label{eq:A}
\end{eqnarray}
where $ \Sigma_{\sigma}^{\rm R}( \omega ) \equiv {\rm Re} \;
\Sigma_{\sigma}( \omega ) $, and $\phi_\sigma (\omega) $ is the phase of
${\cal G}^R_{d, \sigma} (\omega)$.
%
%
 Since $ \Sigma_{\uparrow}^{\rm R}( \epsilon_{\rm F} ) \neq
\Sigma_{\downarrow}^{\rm R}( \epsilon_{\rm F} ) $, an equal spin
occupation, $ n_\uparrow =  n_\downarrow $, is possible only for $
( { \epsilon_{\rm d \uparrow}
 -\epsilon_{\rm F} + \Sigma_{\uparrow}^{\rm R}( \epsilon_{\rm F} )) /
\Gamma_{\uparrow} } = ( { \epsilon_{\rm d \downarrow}
 -\epsilon_{\rm F} + \Sigma_{\downarrow}^{\rm R}( \epsilon_{\rm F} )) /
\Gamma_{\downarrow} } $, which can be obtained only for an
appropriate external magnetic field $ B = B_{\rm comp}$. For the
latter, in the local moment regime ($n\approx 1$) we have
$n_\uparrow = n_\downarrow \approx 0.5$, so that  $\phi_\uparrow =
\phi_\downarrow \approx \pi/2$, which implies that the peaks of
$A_\uparrow$ and $A_\downarrow$ are aligned. Thus, the Friedel sum
rule clarifies why the magnetic field $B_{\rm comp}$ at which the
splitting of the Kondo resonance disappears, coincides with that
for which $n_\uparrow = n_\downarrow$.
For $ B = B_{\rm comp}$, the spin-dependent amplitude $A_{\sigma}(
\epsilon_{\rm F} )$ of Eq.~(\ref{eq:A}) and the conductance
$G_{\sigma} \sim
\Gamma_{\sigma}A_{\sigma}( \epsilon_{\rm F})$
agree  well with the abovementioned NRG
results [Fig.~\ref{fig4}(a),
~\ref{fig4}(b)].

In conclusion, applying the NRG technique, we showed that the
Kondo effect in a QD attached to ferromagnetic leads is in general
suppressed, because the latter induce a spin splitting of the QD
level, which leads to an asymmetry in the occupancy $n_\uparrow
\neq n_\downarrow$. Remarkably, the Kondo effect
  may nevertheless be restored by applying an external magnetic field
  $B_{\rm comp}$, tuned such that the splitting of the Kondo resonance
  is compensated and the condition $n_\uparrow = n_\downarrow$ is
  fulfilled. Although the Kondo resonance is strongly spin polarized,
it then features a locally-screened state,
a spin-independent conductance, and a Kondo temperature which
decreases with increasing spin asymmetry.


 {\it Acknowledgements.} --
We thank R. Bulla, T. Costi, L. Glazman, W. Hofstetter, H.
Imamura, B. Jones, S. Maekawa, A. Rosch, M. Vojta, and Y. Utsumi
for discussions. This work was supported by the DFG under the CFN,
'Spintronics' RT Network of the EC RTN2-2001-00440, Project
PBZ/KBN/044/P03/2001, OTKA T034243 and the Emmy-Noether program.

 {\it Note added.} --
 After submission of our paper,
 a preprint [M. S. Choi et al.
 cond-mat/0305107] studying a similar problem using the NRG technique
 appeared, with conclusions consistent with our's.


\begin{thebibliography}{99}
\bibitem{RKKY_vs_K}
The remarkable full recovery of the Kondo effect should be contrasted to
the competition between Kondo physics and magnetic RKKY interactions in
heavy-fermion systems \cite{jones}.
There, depending on the relation between the relevant energy scales
$ T_{\rm RKKY} $ and $ T_{\rm K} $,
either the local spin is quenched and no magnetic order occurs
(for $ T_{\rm K} > T_{ \rm RKKY} $), or the local molecular field removes spin
degeneracy and suppresses the Kondo effect (for $T_{\rm K} < T_{\rm RKKY} $).
\bibitem{interaction}
We expect that a finite $s$-electron interaction, $U_s > 0$, will
lead only to a slight renormalization of $T_{\rm K}$. Recently, an
Anderson impurity embedded in a correlated host (without magnetic
order) was studied \cite{fulde,hofstetter1}, and an enhancement of
the Kondo scale $T_{\rm K}$  was found for weak $s$-electron
interaction, $ U_s/D < 1$, whereas for intermediate and strong
interactions, $T_{\rm K}$ was found to be reduced.
\bibitem{Stoner}
A lowest-order perturbation calculation shows that for $ |U +
\epsilon_{\rm d}|, \; | \epsilon_{\rm d} | \ll D_{\sigma} $ the
spin splitting of the bottom of the band $ \Delta D $ (for s-bands
we consider no spin splitting for the band top) generates an
effective local magnetic field $B_{\rm{Stoner}} \approx \pi
\Gamma/2 \; \ln (1 - \Delta D/D_{\uparrow} )$.
 To account for this in our model of
Eq.~(\ref{eq:AMf_cont}), one could renormalize $P$ (or else use a
shifted compensation field $B_{\rm{comp}}$ below).
\bibitem{exception}
 For certain special (non-generic) cases,
 the spin up and down level positions and hence also
 the Kondo resonance does not split up in two, even for $P\neq 0$:
 this occurs, for example, at
 the symmetry point $\epsilon_d = - U/2$ of the Anderson model,
 provided the bands are also symmetric around $\varepsilon_F$ (but
 this is unrealistic for ferromagnets).
\bibitem{friedel}
Our NRG confirms  Eq.~(\ref{eq:A}) for arbitrary $P$ and $B$.
\end{thebibliography}
\end{document}